\documentclass[10pt]{article}
\usepackage[cp1251]{inputenc}
\usepackage[english]{babel}
\usepackage{graphicx}
\title{{\bf \Large Cosmological Model of Interacting Phantom  and Yang-Mills Fields}\\
{\normalsize ~~{\bf V.\,K. Shchigolev}\thanks{E-mail:
vkshch@yahoo.com}}\\
{\small {\it Ulyanovsk State University, 42 L. Tolstoy Str.,
Ulyanovsk 432000, Russia}}\\
\vspace{2mm}
\small \begin{quote}{\bf Abstract} --  In this paper, we
consider a model of interacting phantom and Yang-Mills (YM) fields supposing the dilaton-type coupling. Making use of specific solution for YM equation previously found by the author, we obtain simple exact solutions for the accelerated expansion  of FRW  cosmological model. Besides, we derive the induced potentials of phantom field corresponding to some given regimes of expansion.The effective equation of state (EoS) have been reconstructed for all types of the models considered here.
 \\
\vspace{2,5mm}
{\bf PACS numbers}: 98.80.-k, 98.80.Cq, 04.20.Jb, 95.36.+x\\
{\bf Key words}: Cosmological Model, Interaction, Phantom,
Yang-Mills Field, Accelerated Expansion.\\
\end{quote}}
\date{}

\begin{document}

\maketitle \vspace{-2.5cm}
\section{Introduction}

Present accelerated expansion of the universe is well
proved in many papers \cite{C1}-\cite{C8}. In order to explain so
unexpected behavior of our universe, one can modify the
gravitational theory \cite{C9}-\cite{C14}, or construct various
field models of so-called dark energy (DE) which EoS
satisfies $w= p/\rho< -1/3$.
The simplest candidate of DE is the cosmological constant
with fixed EOS  $ w = -1$.  If it is quintessence then $-1 < w <- 1/3$ and if it is phantom
then $ w <-1$. The constant  EOS  $w = -1$  is called phantom divide.
There are some dark energies which can
cross the phantom divide from both sides \cite{C15}.
So far, a large class of scalar-field DE models have
been studied, including tachyon \cite{C16}, ghost condensate
\cite{C17} and quintom \cite{O1}, \cite{C18}, and so forth. In addition, other
proposals on DE include interacting DE models \cite{C19},
braneworld models \cite{C20}, and holographic DE models
\cite{C21}, etc. The quintom scenario of DE is designed to
understand the nature of DE with EoS across -1. The quintom
models of DE differ from the quintessence, phantom and k-essence
and so on in the determination of the cosmological evolution.
It is appropriate mention here that impossibility to realize quintom (or crossing of the phantom divide) in genera k-essence-like models was shown in \cite{C22}.

Another class of DE models is based on the conjecture that a
vector field can be the origin of DE \cite{C23},\cite{C24}. The YM
field can be a kind of candidate for such a vector field \cite{O2}, 
\cite{C25}, \cite{C26}. At the same time, it is well known that a
pure YM field (with its EoS $w = 1/3$) can not provide
accelerated expansion of the Universe, for which $w < -1/3$ is
required. This is a direct consequence of conformal symmetry of
the Lagrangian for a massless YM field. Any violations of
conformal symmetry (e.g., as a result of quantum corrections
\cite{C27} or of non-minimal coupling to gravity \cite{C28}) give
a good chance for involving YM fields in reconstruction of DE. The
alternative chance for YM fields to be involved in DE problem is
consideration of some interaction of YM  field with different sources of gravity.
In this aspect,  the idea of induced nonlinearity
is fairly attractive for realization in cosmology of phantom field.  In this paper, we turn our attention to the
issue of the YM fields interacting with a phantom field in FRW  cosmology.
Using the specific solution of YM equation previously considered in FRW cosmology  \cite{C29}-\cite{C33}, we generalize the model investigated in  \cite{C29} on the case of interacting phantom and YM fields.  This allows us to obtain some exact solutions for the accelerated expansion  of FRW  cosmological model. Besides, we derive the induced potentials of phantom field corresponding to some given regimes of expansion. The effective EoS have been reconstructed for all types of the model considered below.

\section {Basic equations}
The main equations of the model follow from the Lagrangian  density  \cite{C29}:
\begin{equation}
\label{1}
{\cal L}= \frac{R}{{2\kappa} } + \epsilon \frac{{1}}{{2}}\varphi _{,\alpha}
\varphi ^{,\alpha}  - \frac{{1}}{{16\pi} }F_{\alpha \beta} ^{a} F^{a\alpha
\beta} \Psi(\varphi),
\end{equation}
 where R is the Ricci curvature scalar,  $\varphi $  is a scalar field,
 $F_{\alpha \beta} ^{a} = \partial _{\alpha}  W_{\beta} ^{a} - \partial
_{\beta}  W_{\alpha} ^{a} + g\varepsilon _{abc} W_{\alpha} ^{b} W_{\beta
}^{c}$ is the YM strength tensor,
 $\Psi \left( {\varphi}  \right) \quad$ is the coupling analytical function of the phantom and YM fields.  $ \epsilon = + 1$  represents  quintessence while  $ \epsilon = - 1$ refers to phantom field.

As it is noted in \cite{C29}, the similar Lagrangian is well founded by the existence of process $\pi\to 2\gamma$.
The case $\Phi(\varphi)=\exp (-2\lambda \varphi)$ with $\lambda = \sqrt{3}$ occurs via the Kaluza-
Klein compactification of five-dimensional vacuum gravity.
The critical coupling $\lambda = 1$ arises by the truncation
of $N=4$ supergravity.The same type of interaction emerges in Brans-Dicke theory with $\lambda = 1/\sqrt{\displaystyle 2 \omega + 3}$. This fact encourages us to suppose that the theory (\ref{1}) can be embedded into some cosmological theories for general coupling.

Variation of (\ref{1}) with respect to metrics $g^{ik}$  and fields yields
the Einstein equation
\begin{equation}
\label{2}
G_{\mu} ^{\nu}  = \kappa T_{\mu} ^{\nu}  ,
\end{equation}
where $G_{\mu} ^{\nu}  $ is the Einstein tensor, $\kappa$  is the Einstein gravitational constant. From (\ref{1}), one can obtain the following energy-momentum tensor for the system of fields,
\begin{equation}
\hspace{-1cm}T_{\mu} ^{\nu}  = \epsilon \varphi _{,\mu}  \varphi ^{,\nu}
 - \frac{1}{4\pi
}F_{\alpha \beta} ^{a} F^{a\alpha \beta} \Psi (\varphi)-\delta _{\mu} ^{\nu}  \left[ \frac{\epsilon}{2} \varphi _{,\alpha}
\varphi ^{,\alpha}  - \frac{1}{16\pi}F_{\alpha \beta} ^{a} F^{a\alpha
\beta} \Psi (\varphi)\right],
\label{3}
\end{equation}
and the modified YM equation:
\begin{equation}
\label{4}
D_{\nu}  \left(\sqrt{ - g} F^{a\nu \mu} \Psi(\varphi)
\right) = 0,
\end{equation}
where $D_{\nu}  $ denotes the covariant derivative.
The scalar field equation is as follows:
\begin{equation}
\label{5}
\frac{{ \epsilon} }{{\sqrt { - g}} }\frac{{\partial} }{{\partial x^{\nu} }}\left(
{\sqrt { - g} g^{\nu \mu} \frac{{\partial \varphi} }{{\partial x^{\mu} }}}
\right) + \frac{{1}}{{16\pi} }\, I\, \Psi
_{\varphi}  = 0,
\end{equation}
where $\Psi_{\varphi}  =\displaystyle  \frac{d\Psi(\varphi)}{d\varphi
}$, and the first invariant of Yang-Mills field $ I =F_{\alpha \beta} ^{a} F^{a\alpha \beta}$.

We assume that the Universe is described by a
Friedmann-Robertson-Walker (FRW) geometry:
\begin{equation}
\label {6} d s^2 = d t^2- a^2 (t)(d r^2+\xi^2 (r)d \Omega
^2),
\end{equation}
where $\xi (r)=\sin r,r,\sinh r$ for the sign of space curvature
$k=+1,0,-1$, consequently.
As well-known, the generalized Wu-Yang ansatz for the $SO_3$ YM
fields can be written as \cite{C29}
$$W^a_0 = x^a \frac{W(r,t)}{gr},\quad W^a_{\mu} =\varepsilon_{\mu
ab}x^b\frac{K(r,t)-1}{gr^2}+\Bigl(\delta^a_{\mu}-\frac{x^ax_{\mu}}{r^2}
\Bigr)\frac{S(r,t)}{gr}.
$$
We can make the following substitution into this ansatz
\cite{C31}:
$$
W(r,t)=\dot \alpha(t),\quad K(r,t)= P(r)\cos \alpha(t),\quad S(r,t)= P(r)\sin \alpha(t).
$$

As a result, we have the following expressions for the YM strength
tensor components:
\begin{eqnarray}
&{\bf F}_{01}={\bf F}_{02}={\bf F}_{03}=0,
\quad{\bf F}_{12}=g^{-1} P'(r)\Bigl({\bf m }\,\cos\alpha + {\bf
l}\,\sin\alpha\Bigr),\nonumber\\
&{\bf F}_{13}=g^{-1} P'(r)\sin\theta\Bigl({\bf m }\,\sin\alpha-{\bf
l}\,\cos\alpha\Bigr),\nonumber\\
&{\bf F}_{23}=g^{-1}\sin\theta\Bigl(P^2(r)-1\Bigr){\bf n}, \label {7}
\end{eqnarray}
where \,\,$ {\bf n}= (\sin\theta \cos \phi, \sin \theta \sin \phi,
\cos \theta),~~ {\bf l}= (\cos\theta \cos \phi, \cos \theta \sin
\phi, -\sin \theta)$ and ${\bf m}= (-\sin \phi, \cos \phi, 0) $ are
the orthonormalized isoframe vectors, and the prime means a
derivative with respect to $r$. As noted in \cite{C31}, the YM
field (\ref{7}) has only magnetic components. It is easy to find
from (\ref{6}) and (\ref{7}) that the YM field invariant $I =
F^a_{ik}F^{aik}$ becomes as follows:
\begin{equation}
\label {8} I = \frac{2}{g^2 a^4 \xi^2}\Bigl[ 2 P'^2+\frac{(P^2
-1)^2}{\xi^2} \Bigr].
\end{equation}
With the help of (\ref{7}), assuming the spacial homogeneity of phantom field , it is easy to show that YM equation (\ref{4}) reduces
to the following simple equation \cite{C29}:
\begin{equation}
\label {9}  P''-\frac{(P^2 -1)P}{\xi^2}= 0,
\end{equation}
where $P''\equiv d^2 P/d r^2$.

The nontrivial solution for equation (\ref{9}) obtained in
\cite{C29} is as follows:  $P(r)= \xi'(r)=\cos r, \cosh r$ for $k=+1,-1$,
consequently.  From the latter and (\ref{8}), it follows that the valuable feature of this
solution is that the YM invariant built on this solution depends only on
time:
\begin{equation}
\label {10} I = I(t) = \frac{6}{g^2 a^4(t)}.
\end{equation}
In view of (\ref{10}), the rest set of Einstein and scalar field equations is as follows:
\begin{equation}
\label{11}
\frac{3}{a^2}(\dot {a}^{2} + k) = \kappa \epsilon
\frac{\dot {\varphi} ^{2}}{{2}}+ \frac{3\kappa }{{8\pi g^2}}\frac{{1}}{{a^{4}}}\Psi (\varphi),
\end{equation}
\begin{equation}
\label{12}
\frac{1}{a^2}(\dot {a}^{2} + 2a\ddot {a} + k) = - \kappa \epsilon
\frac{\dot {\varphi} ^{2}}{2} - \frac{\kappa}{8\pi g^2}\frac{1}{a^{4}}\Psi(\varphi),
\end{equation}
\begin{equation}
\label{13}
 \epsilon \frac{{1}}{{a^{3}}}\frac{{\partial} }{{\partial t}}\left( {a^{3}\dot
{\varphi} } \right) + \frac{{3}}{{8\pi g^2}}\frac{{1}}{{a^{4}}}\Psi
_{\varphi}  = 0.
\end{equation}

It is easy to verify that only two equations among Eqs. (\ref{11})-(\ref{13}) are independent. That is why we should define one of the three functions $a(t)$, $\varphi(t)$ or
$\Psi(\varphi)$ to solve the system. The most natural way is
to define the coupling function. But it is not necessary
and depends on a specific context of the problem.
In works \cite{C29}, \cite{C31}, this set of equations is investigated from the assumption of some given regimes for expansion, $k = \pm 1$ and $\epsilon=+1$. Now we consider the phantom case, that is $\epsilon=-1$.

\section{ Simple examples of exact solution}
Let us study the set of dynamics equations (\ref{11})-(\ref{13}) for the case of phantom field considering $ \epsilon = -1$. The set of Eqs.(\ref{11})-(\ref{13}) can be re-written as the following two independent equations for $\dot \varphi$ and $\Psi (\varphi)$:
\begin{equation}
\label{14}
\hspace{-1cm}\dot {\varphi} ^{2} = 6\kappa ^{ - 1}a^{ - 2}(
\dot {a}^{2} + a\ddot {a} + k),\quad \Psi(\varphi) = 8\pi g^2\kappa ^{ - 1}a^{2}(2\dot
{a}^{2} + a\ddot {a} + 2k).
\end{equation}
Moreover, comparing Eq. (\ref{13}) in the case $ \epsilon = - 1$ with the usual one for a phantom field in FRW cosmology,
$$
\ddot {\varphi}  + 3H\dot {\varphi} -
\frac{{dV_{eff} }}{{d\varphi} } = 0,
$$
where $H =\displaystyle  \frac{{\dot
{a}}}{{a}}$ is the Hubble parameter, we derive the relationship between an effective potential $V_{eff}$ and the coupling function $\Psi(\varphi)$:
\begin{equation}\label{15}
\frac{{dV_{eff}(\varphi)} }{{d\varphi} } = \frac{{3}}{{8\pi g^2}}\frac{{1}}{{a^{4}}}\frac{d \Psi (\varphi)}{d\varphi}
\end{equation}

From the first equation of (\ref{14}) it follows that the real solution for phantom field exists when
\begin{equation}
\label{16} \dot {a}^{2} + a\ddot {a} + k \ge 0.
\end{equation}
Besides, the simple study of effective energy density and pressure which follow from the r.h.s. of (\ref{11}) and (\ref{12}),
\begin{equation}
\rho(t) =\displaystyle  - \frac{\dot {\varphi}
^{2}}{2} + 3\frac{1}{8\pi g^2a^{4}}\Psi(\varphi),\quad
p(t) =\displaystyle - \frac{\dot {\varphi}
^{2}}{2} + \frac{1}{8\pi g^2a^{4}}\Psi(\varphi), \label{17}
\end{equation}
shows that the effective EoS in this model is restricted by $p \le - \rho/3$. Therefor, the accelerated expansion, for which $p < - \rho/3$ is required, can be realized in our model.

To demonstrate some interesting features of this model, we are going to consider two illustrative examples of exact solutions further to our general study. We will consider the accelerated expansion, that is $\ddot {a} > 0$, when the asymptotic value of the Hubble parameter $H=H_0 = const$ is achieved in the course of time. For all that, the necessary condition (\ref{16}) is satisfied at every instant.

\subsection{The case ~~{\normalsize $a(t) = H_{0}^{ - 1}\sinh (H_{0} t)$} }
 Note that this dependence of $a(t)$ on time is satisfies inequality (\ref{16}) for all signs of the curvature, and Eq.(\ref{14}) becomes as follows:
\begin{equation}
\dot \varphi ^2 = \frac{6H_{0}^{2}}{\kappa\sinh^2 (H_{0} t)} \Big(
{2\sinh^{2}(H_{0} t) +1 + k} \Big), \label{18}
\end{equation}
\begin{equation}
\Psi(t) =\! \frac{8\pi g^2\sinh^{2}(H_{0} t)}
{\kappa H_{0}^{2}}\Big( {3\sinh^{2}(H_{0} t) +2\! +\! 2k} \Big). \label{19}
\end{equation}
Consider the cases of negative and positive signs of curvature separately. Here and further, we use the following notations:
$$
\lambda = \sqrt{\frac{\kappa}{12}},\quad \Psi_0 = \frac{24\pi g^2}{\kappa H_0^2},\quad V_0=\frac{3 H_0^2}{\lambda^2}.
$$
\noindent i) Open model: $k= -1$.
From Eqs. (\ref{18}), (\ref{19}), one can find that
\begin{equation}
\label{20} \dot \varphi  = \lambda ^{ - 1}H_{0},\quad \Psi(t) = \Psi_0
\sinh^{4}(H_{0} t).
\end{equation}
Integrating the first equation and taking into account the second one together with the explicit expression of $a(t)$, we obtain that:
\begin{equation}
\varphi(t) = \lambda ^{- 1}H_{0} t +
\varphi _{0},\quad\Psi(\varphi)= \Psi_0
\sinh^{4}\Big[\lambda(\varphi-\varphi_0)\Big]. \label{21}
\end{equation}
With the help of equation (\ref{15}), we can derive the following effective potential which is induced by the coupling to the YM field:
\begin{equation}
\label{22} V_{eff}(\varphi) = V_0 \ln\Big(\sinh \Big[
\lambda (\varphi - \varphi _{0})\Big]\Big)+U_0,
\end{equation}
where $U_0$ is a constant.

\noindent ii) Closed model: $k=+1$.
As it follows from Eqs. (\ref{18}), (\ref{19}) in this case, we have
\begin{equation}
\label{23}
\hspace{-1cm}\dot \varphi=\lambda^{-1}H_0 \coth (H_0 t),\quad\Psi(t) = \Psi_0 \sinh^{2}(H_{0} t)\Big( {\sinh^{2}(H_{0}t)+\frac{4}{3}}
\Big).
\end{equation}
Then from Eq. (\ref{23}), we can find that
\begin{eqnarray}
\varphi(t) = \lambda ^{ - 1}\ln\Big[\sinh (H_{0} t)\Big] + \varphi
_{0},\nonumber\\
\Psi (\varphi)=\Psi_0 e^{\displaystyle 2\lambda(\varphi-\varphi_0)}\Big[e^{\displaystyle 2\lambda(\varphi-\varphi_0)}+\frac{4}{3} \Big]. \label{24}
\end{eqnarray}
The induced phantom potential, which corresponds to the coupling function (\ref{24}), can be obtained in the following form:
\begin{equation}
\label{25}
V_{eff}(\varphi) = V_0
\Big[ \lambda (\varphi - \varphi _{0}) -\frac{1}{3} e^{\displaystyle -2\lambda(\varphi-\varphi_0)}\Big]+U_0.
\end{equation}

\subsection{The case ~~$a(t) = H_{0}^{ -
1}\cosh (H_{0} t)$}
In this case, Eqs. (\ref{14}) take the following form:
\begin{equation}
\dot {\varphi} ^{2} = \frac{6H_{0}^2}{\kappa  \cosh^2 (H_{0} t)}\Big(
{2 \cosh^{2}(H_{0} t)-1 + k} \Big), \label{26}
\end{equation}
\begin{equation}
\label{27}
\Psi(t) =\frac{8\pi g^2\cosh^{2}(H_{0} t)}{\kappa H_0^2
}\Big( {3\cosh^{2}(H_{0} t)\! -2 + 2k} \Big),
 \end{equation}

\noindent iii) Open model: $k= -1$.
From Eqs. (\ref{26}), (\ref{27}) we have
\begin{equation}
\hspace{-1cm}\dot \phi = \lambda^{-1}H_0 \tanh (H_0 t),\quad \Psi(t) = \Psi_0\cosh^2(H_{0} t)\Big( {\cosh^{2}(H_{0} t) - \frac{4}{3}} \Big). \label{28}
\end{equation}
In view of the second equation in (\ref{28}) and explicit expression for $a(t)$, we can solve the first equation in (\ref{28}), and then obtain
\begin{eqnarray}
\varphi(t) = \lambda ^{ - 1}\ln \Big[\cosh (H_{0} t)\Big] + \varphi
_{0},\nonumber\\
\Psi (\varphi)=\Psi_0 e^{\displaystyle 2\lambda(\varphi-\varphi_0)}\Big[e^{\displaystyle 2\lambda(\varphi-\varphi_0)}-\frac{4}{3} \Big].\label{29}
\end{eqnarray}
Due to Eq. (\ref{15}), the effective phantom potential becomes
\begin{equation}
\label{30}
V_{eff}(\varphi) = V_0
\Big[ \lambda (\varphi - \varphi _{0}) +\frac{1}{3} e^{\displaystyle -2\lambda(\varphi-\varphi_0)}\Big]+U_0.
\end{equation}
\begin{figure}[t]
\centering
\includegraphics[width=90mm,height=6cm]{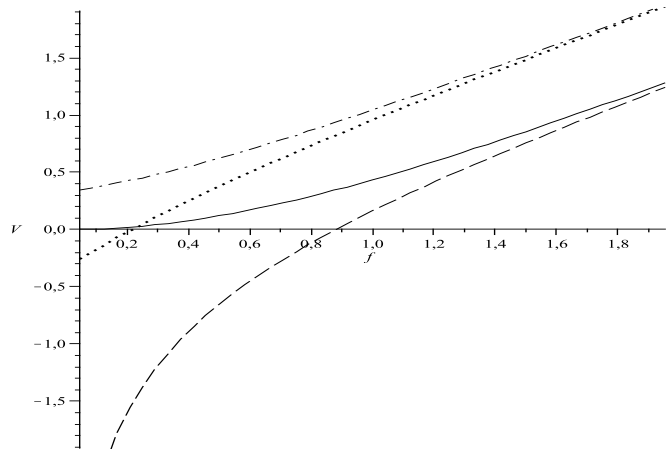}\\
\caption{Effective potential $V_{eff}$ ($V_0=1, U_0=0$) versus $f=\lambda\phi$ for $a=H_0^{-1}\sinh(H_0t)$
when $k=-1$ (dashed curve) or $k=+1$ (dotted curve),
and for $a=H_0^{-1}\cosh(H_0t)$ when $k=-1$ (chain curve) or $k=+1$ (solid curve).}
\label{pic1}
\end{figure}
\noindent iv) Closed model: $k=+1$.
For this sign of curvature, it is follows from Eqs. (\ref{26}), (\ref{27}) that
\begin{equation}\label{31}
 \dot \varphi  = \lambda ^{ - 1}H_{0},\quad  \Psi(t) = \Psi_0
\cosh^{4}(H_{0} t).
\end{equation}
Therefor, we have the following solution for (\ref{31}):
\begin{equation}
\varphi \left( {t} \right) = \lambda ^{ - 1}H_{0} t +
\varphi _{0},\quad\Psi(\varphi)= \Psi_0
\cosh^{4}\Big[\lambda(\varphi-\varphi_0)\Big]. \label{32}
\end{equation}
In view of (\ref{15}), it can be easily obtained that now the effective phantom potential is
\begin{equation}
V_{eff}(\varphi) = V_0\ln\Big(\cosh\Big[\lambda
(\varphi - \varphi _{0})\Big]\Big)+U_0. \label{33}
\end{equation}
The plots of effective potentials for all cases considered can be viewed on Fig.1.

With the  help of expressions (\ref{17}), all solutions obtained above can be arranged in tree groups according to their EoS.

In case (1): $a(t) = H_{0}^{ -
1}\cosh (H_{0} t)$, ($k=-1$), and in case (2): $a(t) = H_{0}^{ -
1}\sinh (H_{0} t)$, ($k=+1$), one can find consequently
\begin{equation}\label{34}
w_{1} = -\frac{3 \sinh ^2 (H_0 t)+1}{3 \sinh ^2 (H_0 t)-3},\quad w_{2} = -\frac{3 \sinh ^2 (H_0 t)+2}{3 \sinh ^2 (H_0 t)+6}
\end{equation}
In case (3): $a(t) = H_{0}^{ -
1}\sinh (H_{0} t)$ with $k=-1$, or  $a(t) = H_{0}^{ -
1}\cosh (H_{0} t)$ with $k=+1$, we have  $w_{3}=-1$.
The plots of EoS parameters for all three cases considered above can be viewed on Fig.2.

As one can see, the curve of EoS in the case $a=H_0^{-1}\sinh(H_0t)$
with $k=-1$ displays rather unusual behavior around the value $x=H_0t$ determined by the solution of $\sinh(H_0t)=1$. To understand this, let us find the energy densities from Eq. (\ref{17}) for the cases defined above. So we can find for the cases (1) and (2) that
\begin{equation}\label{35}
\rho_1= \frac{H_0^2}{4 \lambda^2 \cosh^2(H_0 t)}\Big(\sinh^2(H_0 t)-1\Big),
\end{equation}
\begin{equation}\label{36}
\rho_2= \frac{H_0^2}{4 \lambda^2 \sinh^2(H_0 t)}\Big(\sinh^2(H_0 t)+2\Big).
\end{equation}
For case (3), we have $\displaystyle \rho_3=\frac{H_0^2}{4 \lambda^2}$, that is $\displaystyle \rho_{3} = \lim_{t \to \infty} \rho_1(t)=\lim_{t \to \infty} \rho_2(t)$. It could be supposed that the constant density $\rho_3$ represents some energy density of an effective cosmological constant $\Lambda$, that is $\displaystyle \frac{\Lambda}{\kappa}=\frac{H_0^2}{4 \lambda^2}$. This implies the well-known result for a de Sitter model: $H_0=\sqrt{\Lambda/3}$. At the same time, one can conclude from (\ref{35}) that while $\sinh^2(H_0 t)<1$ the energy density $\rho_1$ is negative, i.e. the weak energy condition is violated. But as soon as $\sinh^2(H_0 t)>1$, $\rho_1$ becomes positive, and the energy condition takes effect. From  (\ref{36}), it is obvious that there is no any problem of this sort in case (2), as well as in case (3).
\begin{figure}[t]
\centering
\includegraphics[width=90mm,height=6cm]{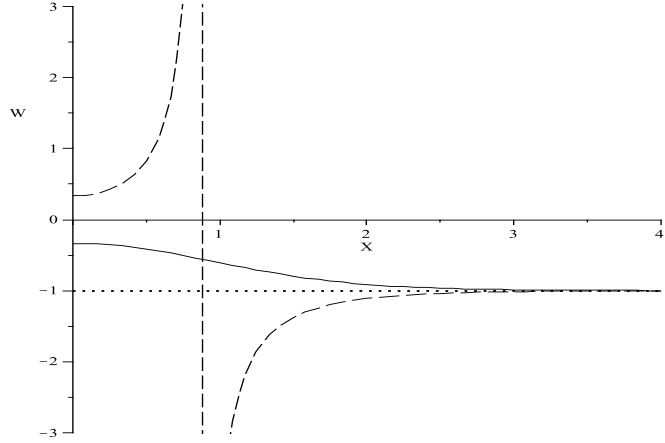}\\
\caption{Effective EoS parameter  $w$ versus $x=H_0 t$ for $a=H_0^{-1}\sinh(H_0t)$
when $k=-1$ (dotted curve) or $k=+1$ (solid curve),
and for $a=H_0^{-1}\cosh(H_0t)$ when $k=-1$ (dashed curve) or $k=+1$ (dotted curve).}
\label{pic2}
\end{figure}

\section{ Conclusion}
In summary, the model of interacting phantom and YM fields in FRW
non-flat cosmology are shortly studied in this paper. First of all, we
have derived the set of main equations which determines the model
dynamics: (\ref{11}), (\ref{12}) and (\ref{13}). Making use of the specific solution of YM equation previously considered in FRW scalar field cosmology, we generalized the model investigated in  \cite{C27} on the case of interacting phantom and YM fields.  This allowed us to obtain some exact  solutions for the accelerated expansion  of FRW  cosmological model. Besides, we derive the induced potentials of phantom field corresponding to the cases (3.1) and (3.2) in which the Hubble parameter changes as $H(t) = H_{0}\coth ( H_{0} t)$ or $H(t) = H_{0} \tanh (H_{0} t)$ consequently. At that, it is follows from (\ref{20}), (\ref{23}), (\ref{28}) and (\ref{31}) that all cases considered above are related by the conditions $\dot {\varphi} \sim H_{0} $ or $\dot {\varphi}
\sim H(t)$. In other words, the rate of phantom field change  is proportional to either asymptotical value of the Hubble parameter ( $\displaystyle  H_{0} = \lim_{t \to \infty}  H(t)$ ) or to its contemporary value.
The effective EoS have been reconstructed for all types of the models considered above. Somewhat unexpected is the results of cases (1). Nevertheless, this model considered from the moment $t=H_0^{-1} \sinh^{-1}(1)$ does not demonstrate any oddity. As can be seen from our examples, all EoS parameters considered do not cross the phantom divide $-1$ at late time. This is not a characteristic property of the model but only the consequence of the simplest expansion  regimes given.


\begin{thebibliography}{99}
{\small
\bibitem{C1} S. Perlmutter et al., Astrophys. J. 517, 565 (1999).
\vspace{-2mm}
\bibitem{C2} C. B. Netterfield et al., Astrophys. J. 571, 604 (2002).
\vspace{-2mm}
\bibitem{C3} N.\,W. Halverson et al., Astrophys. J. 568, 38 (2002).
\vspace{-2mm}
\bibitem{C4} S. Bridle, O. Lahab, J. P. Ostriker and P. J. Steinhardt, Science 299, 1532
(2003).
\vspace{-2mm}
\bibitem{C5} D.\,N. Spergel et al., Astrophys. J. Suppl.
Ser. 148, 175 (2003).
\vspace{-2mm}
\bibitem{C6} C.\,L. Bennett, et al., Astrophys. J. Suppl. {\bf 148},
1 (2003).
\vspace{-2mm}
\bibitem{C7} M. Tegmark, et al.[SDSS Collaboration], Phys. Rev. D
{\bf 69}, 103501 (2004)
\vspace{-2mm}
\bibitem{C8} S.\,W. Allen, et al.,Mon. Not. Roy. Astron. Soc. {\bf 353}
457 (2004).
\vspace{-2mm}
\bibitem{C9} G.\,R. Dvali,
G. Gabadadze, M. Porrati, Phys. Lett. B 485, 208 (2000).
\vspace{-2mm}
\bibitem{C10} S. Nojiri and S.\,D. Odintsov,  Phys. Rev. D 68,
123512 (2003).
\vspace{-2mm}
\bibitem{C11} S. Capozziello, Int. J. Mod. Phys. D 11, 483
(2002).
\vspace{-2mm}
\bibitem{C12} P.\,S. Apostolopoulos, et al., Phys. Rev. D 72, 044013 (2005).
\vspace{-2mm}
\bibitem{C13} S. Nojiri and S.\,D. Odintsov, Int. J. Geom.
Meth. Mod. Phys. 4, 115 (2007).
\vspace{-2mm}
\bibitem{C14} F.\,K. Diakonos and E.\,N. Saridakis, JCAP 0902,
030 (2009).
\vspace{-2mm}
\bibitem{C15} C. Deffayet, O. Pujolas, I. Sawicki, A. Vikman, JCAP 1010,
026 (2010).
\vspace{-2mm}
\bibitem{C16} A. Sen, JHEP {\bf0207}, 065 (2002).
\vspace{-2mm}
\bibitem{C17} F. Piazza, S. Tsujikawa, JCAP {\bf 0407}, 004 (2004).
\vspace{-2mm}
\bibitem{O1} E. Elizalde, S. Nojiri and S.\, D. Odintsov, hep-th/0405034.
\vspace{-2mm}
\bibitem{C18} B. Feng, X. Wang and X. Zhang, Phys. Lett. B {\bf 607}, 35
(2005).
\vspace{-2mm}
\bibitem{C19} M.\,R. Setare, Phys. Lett. B {\bf 642}, 1 (2006).
\vspace{-2mm}
\bibitem{C20} E.\,N. Saridakis, Phys. Lett. B {\bf 660}, 138 (2008).
\vspace{-2mm}
\bibitem{C21} E. Elizalde, S. Nojiri, S.\,D. Odintsov and P. Wang, Phys. Rev. D {\bf 71}, 103504 (2005).
\vspace{-2mm}
\bibitem{C22}  A. Vikman , Phys. Rev. D{\bf 71}, 023515 (2005).
\vspace{-2mm}
\bibitem{C23} V.\,V. Kiselev, Class. Quant. Grav.,{\bf 21}, 3323 (2004).
\vspace{-2mm}
\bibitem{C24} H. Wei and R.\,G. Cai, Phys. Rev. D {\bf 73}, 083002 (2006).
\vspace{-2mm}
\bibitem{O2} E. Elizalde, J.\, E. Lidsey, S. Nojiri and S.\, D. Odintsov, hep-th/0307177.
\vspace{-2mm}
\bibitem{C25} Y. Zhang, Phys.Lett. B {\bf 340}, 18 (1994).
\vspace{-2mm}
\bibitem{C26} Y. Zhang, T.\,Y. Xia and W. Zhao, Class. Quant. Grav.,{\bf 24}, 3309 (2007).
\vspace{-2mm}
\bibitem{C27} W. Zhao, Y. Zhang, Class. Quant. Grav., {\bf 23} 3405 (2006).
\vspace{-2mm}
\bibitem{C28} D.\,V. Gal'tsov, arXiv: 0901.0115 [gr-qc] (2009).
\vspace{-2mm}
\bibitem{C29} V.\,K. Shchigolev and M.\,V. Shchigolev,  J. Exp. Theor. Phys., Vol. 119, No.4, 1 (2001).
\vspace{-2mm}
\bibitem{C30} V.\,K. Shchigolev, S.\,V. Chervon, O.\,V. Kudasova,  Grav. Cosmol., Vol. 32, No.1, 41 (2000).
\vspace{-2mm}
\bibitem{C31} V.\,K. Shchigolev, K. Samaroo, Gen. Relat. Grav., Vol. 36, No. 7, 1661 (2004).
\vspace{-2mm}
\bibitem{C32} V.\,K. Shchigolev, Grav. Cosmol., Vol. 17, No.3, 272 (2011).
\vspace{-2mm}
\bibitem{C33} V.\,K. Shchigolev, G.\,N. Orekhova,  Mod. Phys. Lett. A., Vol. 26, No. 26, 1965 (2011).}
\end{thebibliography}
\end{document}